\documentstyle[multicol,aps,prl,psfig]{revtex}

\begin{document}

\title{Chain Collapse and Counterion Condensation
in Dilute Polyelectrolyte Solutions}

\author{ N.~V.~Brilliantov$^{1,2,3}\,$, D.~V.~Kuznetsov$^{4}$, 
and  R.~Klein$^3$}
\address{$^{(1)}$Department of Chemistry, University
of Toronto, Toronto, Canada M5S 1A1}
\address{$^{(2)}$Moscow State University, Physics Department,
Moscow 119899, Russia}
\address{$^{(3)}$ Universitat Konstanz, Fakultat fur Physik,
Universitatsstrasse 10, Postf. 5560 M671,D-78434, Konstanz, Germany}
\address{$^{(4)}$ Institute of Biochemical Physics, Russian Academy
of Sciences, Kosygin St. 4, Moscow 117977, Russia}

\maketitle

\begin{abstract}

A new quantitative theory for polyelectrolytes in salt free dilute
solutions is developed. Depending on the electrostatic interaction
strength, polyelectrolytes in solutions can undergo strong stretching (with
polyelectrolyte dimension $R_g\sim l_B^{1/3}N$, where $l_B$ is the Bjerrum
length and $N$ is the number of the chain segments) or strong
compression (with $R_g\sim l_B^{-1/2}N^{1/3}$). A strong polymer collapse
occurs as a first-order phase transition due
to accompanying counterion condensation.

\end{abstract}

\begin{multicols}{2}

Many important synthetic and biological macromolecules are polyelectrolytes
and their properties differ significantly from that  of the neutral polymers
\cite{GrKh}. Importance of the former stimulates a variety of their
analytical \cite{Barr} and numerical \cite{Kremer1} studies.
However, even for a simplest system of a salt-free solution of linear
flexible polyelectrolytes (an electroneutral system of charged chains and
counterions), many fundamental properties are still unclear or described on
a phenomenological level only.

The neutral chains undergo a collapse transition as the solvent quality
decreases \cite{GrKh,K1}, but collapse and stretching of polyelectrolytes
follow quite different laws. This happens due to a particular role of the
counterions in these processes, which can not be reduced only to a simple 
electrostatic screening and to an increase of the persistence length
\cite{Odijk}. Under conditions of extreme dilution, for a weakly-charged
polyelectrolytes, the counterions occupy the whole volume almost 
uniformly, with a very low concentration, owing to the entropy ``forces''.
Under these conditions, the polyelectrolytes tend to strong stretching
caused by the strong (weakly screened) inter-segment repulsions; the chain
dimensions in this case are proportional to a number of charged segments.
Contrary, under a strong polyelectrolyte charge, when the electrostatic
energy of polymer-counterion attraction is larger than the corresponding
loss of entropy due to the counterion localization, an essential part of
the counterions localizes in the close vicinity of the polyelectrolytes.
This effect, called the counterion condensation \cite{Mann}, leads to an
effective polymer charge screening and can cause an essential decrease of
the polymer size. The corresponding chain collapse can be even stronger
than for the well-studied neutral chains.

In this letter we present a new quantitative theory of the polyelectrolytes
in salt free dilute solutions. We describe the polyelectrolyte dimension in 
a wide range of parameters, covering the area of  
the counterion condensation,  and analyze the nature
of the condensation.

Consider a dilute solution (electroneutral as a whole) of charged chains,
each composed of $N\gg 1$ segments (monomers)
 with the bond-length  $a$. Let each
monomer have a unit charge $e$ (say positive) so that $N$ counterions of
the opposite charge (say negative) are present. Let $V=(4\pi/3)R_{ws}^3$ be
the volume per one chain with the $R_{ws}$ being the Wigner-Seitz radius,
and we assume that the solution is very dilute, so that the conditions $R_g
\ll R_{ws}$ and $r_D \ll R_{ws}$ hold. Here $R_g$ is the gyration radius of
the polyelectrolyte chain and $r_D$ is the Debye screening length built up
on the counterion subsystem. Thus, we consider the case when the counterion
clouds of the different polyelectrolyte chain do not overlap so that we can
concentrate on the one-chain problem. The Bjerrum length in the system is
$l_B=\beta e^2/ \varepsilon$, where $\varepsilon$ is the dielectric
permittivity of the solvent, $\beta\equiv 1/(k_B T)$ with $T$ and $k_B$
being the temperature and the Boltzman factor, respectively. We also assume
that the point-counterions approximation may be used.

We use the following Hamiltonian for the one-chain problem of interest:
%
%%%%%%%%%%     HAMILTONIAN        %%%%%%%%%%%%
\begin{equation}
\label{ham1}
H = H_{\rm n.ch} + H_{\rm el.ch}
+ H_{\rm c}
+ H_{\rm c-ch} \ .
\end{equation}
Here $H_{\rm n.ch}$ is the Hamiltonian of the neutral chain, which accounts
for all noncoulombic interactions between the monomers of the chain,
$H_{\rm el.ch}$ accounts for the Coulombic interactions between 
the monomers of
the chain, $H_{\rm c}$ contains the ideal counterion part and the Coulombic
part of the counterion-counterion interactions; finally, $H_{\rm c-ch}$
accounts for the Coulombic interactions between the counterions and the
chain. It is convenient to write the Coulombic interactions in terms of the
microscopic densities of the counterions,
$\hat{\rho}_c \left({\bf r}\right)
=\sum\limits_{i=1}^{N} \delta \left({\bf r}-{\bf r}_{i} \right)$,
and monomers,
$\hat{\rho}_m ({\bf r}) =\sum\limits_{j=1}^{N} \delta \left({\bf r}
- {\bf R}_j \right) $,
where $\left\{ {\bf r}_{i} \right\}$ and $\{ {\bf R}_{j} \}$ are 
coordinates of the counterions and of the 
monomers respectively. $H_{\rm el.ch}$
and $H_{\rm c-ch}$ read in this notations \cite{remark11}:
%
%%%%%%%%%%%%%  COULOMB INTERACTIONS  H_el.ch & H_c-ch  %%%%%%%%%%%
\begin{eqnarray}
\label{ucoul}
&&\beta H_{\rm el.ch}
=\frac{l_B}{2} \int d {\bf r} d {\bf r}' \varphi \left( {\bf r}- {\bf r}'
\right)\hat{\rho}_m
\left({\bf r} \right)
\hat{\rho}_m
\left({\bf r}' \right) \ , \\
&& \beta H_{\rm c-ch}= -l_B
 \int d {\bf r} d {\bf r}' \varphi \left( {\bf r}- {\bf r}'
\right)\hat{\rho}_m
\left({\bf r} \right)
\hat{\rho}_c
\left({\bf r}' \right) \ ,
\end{eqnarray}
where $\left(e^2/\varepsilon \right)\varphi (r)=\left(e^2/\varepsilon r
\right)$ is the Coulomb potential. For the subsequent analysis it is 
worth to
map the counterion part, $H_{\rm c}$ onto the Hamiltonian of the
one-component-plasma (OCP) \cite{lowen}. The OCP model is formulated as
follows: the point charges are immersed into the structureless compensating
background of the opposite charge. The background charge density is
$e\rho({\bf r})$, and the average local density of the point charges
(counterions in what follows) is equal to $\rho({\bf r})$ \cite{lowen}. The
OCP Hamiltonian reads:
%
%%%%%%%%%%    HAMILTONIAN OF THE OCP- MODEL  %%%%%%%%%%%
\begin{equation}
\label{hocp}
H_{\rm OCP}\left[ \rho({\bf r}) \right] =
H_{\rm c}+H_{\rm bb}\left[ \rho({\bf r}) \right]
+H_{\rm bc}\left[ \rho({\bf r}) \right] \ ,
\end{equation}
where $\beta H_{\rm bb}= \left( l_B /2\right) \int d {\bf r} d {\bf r}'
\varphi ( {\bf r}- {\bf r}' ){\rho} ({\bf r} )
{\rho} ({\bf r}')$ describes the background self-interaction, while
$\beta H_{\rm bc}= -l_B \int d {\bf r} d {\bf r}' \varphi
( {\bf r}- {\bf r}' ){\rho} ({\bf r} ) \hat{\rho}_c ({\bf r}' )$
gives the energy of the background-counterion interaction. From Eq.
(\ref{hocp}) it follows that $H_{\rm c}=H_{\rm OCP}-H_{\rm bc}-H_{\rm bb}$.

To address the problem of the gyration radius of the chain, we consider the
{\it conditional} free energy of the system, $F(R_g)$, as a function of the
gyration radius $R_g$. It is convenient to define an effective Hamiltonian
of the neutral chain, $\overline{H}_{\rm n.ch}$, as $\exp\left(-\beta
\overline{H}_{\rm n.ch}\right) \equiv D(R_g) \exp\left(-\beta H_{\rm
n.ch}\right)$, where the conditional function $D(R_g)$ is equal to unity if
the coordinates of the monomers are consistent with the condition for the
gyration radius to be $R_g$, otherwise it equals zero. Since the free
energy and structural properties of the systems with Hamiltonians
$\overline{H}_{\rm n.ch}$, and $H_{\rm OCP}$ are known (to some extent), it
is reasonable to write the total Hamiltonian of the system as a sum of two
parts: the reference part,
%%%%%%%%%%%%%%%%%%%%%%%%%%     H_0   %%%%%%%%%%%%%%%%%%%%%%%%%%
\begin{equation}
\label{href}
H_0= \overline{H}_{\rm n.ch} \left(R_g \right) +
H_{\rm OCP}\left[ \rho({\bf r}) \right] \ ,
\end{equation}
and the perturbation part,
\begin{eqnarray}
\label{hper}
&&\beta H' =\frac{l_B}{2} \int d {\bf r} d {\bf r}' \varphi ({\bf r}-{\bf r}')
\left[ \hat{\rho}_m ({\bf r}) \hat{\rho}_m ({\bf r}')
-\rho ({\bf r} ) \rho ({\bf r}' ) \right]
\nonumber \\
&&- l_B \int d {\bf r} d {\bf r}' \varphi ({\bf r}-{\bf r}')
\left[ \hat{\rho}_m ({\bf r}) -\rho ({\bf r} ) \right]
\hat{\rho}_c ({\bf r}')  \ ,
\end{eqnarray}
and use then the Gibbs-Bogoljubov inequality:
%%%%%%%%%%     GIBBS-BOGOLJUBOV INEQUALITY   %%%%%%%%%

\begin{equation}
\label{gibbs}
F(R_g) \le F_{\rm n.ch} (R_g) + F_{\rm OCP}[\rho ({\bf r}) ] +
{\langle H' \rangle}_{H_0} \ .
\end{equation}
Here $F_{\rm n.ch} (R_g)$ is the free energy of the neutral chain with the
gyration radius $R_g$. $F_{\rm OCP}[\rho ({\bf r}) ]$ is the free energy of
the inhomogeneous OCP with the background charge density $\rho ({\bf r})$.
Finally, ${\langle H' \rangle}_{H_0}$ is obtained by averaging the
perturbation part, $H'$, given by Eq. (\ref{hper}) over the reference
Hamiltonian, $H_0$, given by Eq. (\ref{href}). Minimizing the right-hand
side of Eq. (\ref{gibbs}) with respect to $\rho ({\bf r})$, which is equal
to the average density of the counterions, one obtains an estimate to the
conditional free energy, $F(R_g)$. Minimizing then $F(R_g)$ with respect to
$R_g$, one finds the gyration radius.

The free energy of the neutral polymer is a sum of the elastic part,
written in the Flory-type approximation \cite{GrKh} as $k_BT \gamma \left(
\alpha^2 + \alpha^{-2} \right)$ and the interaction part, written on the
level of the second virial approximation \cite{GrKh} as $k_BT BN^2 / \left(
4\pi R_g^3 /3 \right)$. Here $\alpha$ is the chain expansion factor,
$\alpha^2 \equiv R_g^2/R_{g.{\rm id}}^2$, with $R_{g.{\rm id}}$ being the
mean-square gyration radius of the ideal chain, $R_{g.{\rm id}}^2=Na^2/6$,
$\ \gamma \simeq 9/4$ for the Gaussian polymers \cite{K1}, and $B$ is the
second virial coefficient. Here we consider the case of a good solvent,
$B>0$. With the reduced coefficient ${B}^* \equiv 6^{3/2}B/ \left( 3\pi
a^3 \right)$, we write for $F_{\rm n.ch} (R_g)$:
%
%%%%%%%%%   FREE ENERGY OF THE UNCHARGED CHAIN   %%%%%
\begin{equation}
\label{french}
\beta F_{\rm n.ch} (R_g) \simeq \frac94 \left( \alpha^2 + \alpha^{-2}
+ {B}^* N^{1/2} \alpha^{-3} \right) \ . 
\end{equation}

The OCP-part of the free energy reads:
%%%%%%%%%%%%%%%   FREE ENERGY OF THE OCP  %%%%%%%%%%
\begin{eqnarray}
\label{frenocp}
&&\beta F_{\rm OCP}[\rho ({\bf r}) ]=
 \int d {\bf r} \rho ({\bf r}) \ln \left[ \Lambda_c^3
\rho ({\bf r}) -1 \right] \nonumber \\
&&+\int d {\bf r} \rho ({\bf r}) \Psi_{\rm OCP}^{exc} [\rho ({\bf r}) ] \ ,
\end{eqnarray}
where the first term in the right-hand side of Eq. (\ref{frenocp}) refers
to the ideal part of the OCP-free energy ($\Lambda_c$ is the thermal
wavelength of the counterions), while the second term is the excess part,
written in the local density approximation \cite{lowen}. The function
$\Psi_{\rm OCP}^{exc} [ \rho ({\bf r})]$ denotes the excess free energy per
ion in the inhomogeneous OCP-model \cite{lowen}. It may be expressed in
terms of the (local) ``plasma'' parameter $\Gamma=l_B/a_c$, where $a_c=\{
3/[4\pi \rho({\bf r})]\}^{1/3}$ is the (local) ion-sphere radius of the
counterion. For $\Gamma \ll 1 $ ( the Debye-Huckel limit) $\Psi_{\rm
OCP}=-\sqrt{3} \Gamma^{3/2}$, while for $\Gamma \gg 1$ an analytical fit
for the Monte Carlo (MC) data is available \cite{witt}. One can also use an
approximate equation of state (EOS) for the OCP, which covers all the range
of $\Gamma$ \cite{nord}. It gives a reasonable $10\%$ accuracy in the range
of $0 < \Gamma <100$, but loses considerably its accuracy for $\Gamma > 100$.
Recently we proposed much more accurate approximate EOS for the OCP
\cite{brilocp}:
\begin{equation}
\label{Psi_OCP}
\Psi_{\rm OCP}(\Gamma) \equiv \frac34 \left[ \ln (1+c\Gamma)-c \Gamma \right]
- \frac32 (c\Gamma)^{3/2}\arctan\frac{1}{\sqrt{c\Gamma}} \ ,
\end{equation}
where $c=(2/3)\left(2/\pi^2\right)^{1/3}$. This EOS has the correct
Debye-Huckel limit. It agrees within $1-2.5\%$ with the MC data for the
most of the range of $\Gamma$, and has a maximal deviation from the
MC data  (about  $ 10 \%$ ) in the interval $0.1 < \Gamma < 0.5$.
So we use Eq.(\ref{Psi_OCP}) here.

Finally, taking into account that
${\langle \hat{\rho}_c ({\bf r}) \rangle}_{H_0}=\rho ({\bf r})$,
we write for the perturbation part:
%%%%%%%%%%%%%%%%   PERTURBATION   %%%%%%%%%%%%%%
\begin{eqnarray}
\label{perturb}
&&{\beta \langle H' \rangle}_{H_0}
= \frac{l_B}{2} \int d {\bf r} d {\bf r}' \left[ \varphi
\left( {\bf r}- {\bf r'}
\right) g_2
\left({\bf r},{\bf r}' \right) \right.  \\
&&\left. -2 \varphi
\left( {\bf r}- {\bf r}'
\right)g_1
\left({\bf r} \right)
{\rho}
\left({\bf r}' \right)
+ \varphi
\left( {\bf r}- {\bf r}'
\right)
{\rho}
\left({\bf r} \right)
{\rho}
\left({\bf r}' \right) \right] \ , \nonumber
\end{eqnarray}
where $g_2 \left({\bf r},{\bf r}' \right)=
{\langle {\hat{\rho}}_m \left({\bf r} \right)\hat{{\rho}}_m
\left({\bf r}' \right) \rangle}_{H_0}$
is the pair correlation function for the monomer-monomer density inside
the macroion's core (i.e. in the volume confined by the radius of gyration)
and  $g_1 \left({\bf r}\right) = {\langle \hat{{\rho}}_m \left({\bf r} \right)
\rangle}_{H_0}$ is the average monomer density inside the core.

For the counterion density distribution we adopt here a simplified model.
We introduce two characteristic densities: $\rho_{\rm in}$, the average
counterion density inside the macroion's core, and $\rho_{\rm out}$, that
for the outer region.
%We write for the function $\rho(r)$:
%\begin{equation}
%\rho(r) =\rho_{\rm in} \theta (R_g-r) +\rho_{\rm out} \theta (r-R_g) \ ,
%\end{equation}
%where $\theta (x)$ is a unit step function.
This approximation implies that the size of the transient region where the
density of the counterions changes from its in-core value to the bulk value
is small compared to the radius of gyration.
From the normalization condition,
$(4\pi/3) \left( R_{ws}^3-R_g^3 \right) \rho_{\rm out}+
(4\pi/3) R_g^3 \rho_{\rm in} =N$, we deduce that $\rho_{\rm out}$
is determined by $\rho_{\rm in}$. It is convenient to use a dimensionless
density, $\rho \equiv\rho_{\rm in}/n$, where $n=N/V_g$ is the average
monomer density inside the core and $V_g=(4\pi/3) R_g^3$ is the gyration
volume. The monomer correlation functions are approximated as
$g_1 \approx n$, and $g_2 \approx n\cdot n\,=n^2$.

Using these approximations one can find all contributions to the
total free energy. For example, the perturbation part reads:
%%%%%%%%%%  FREE ENERGY, PERTURBATION  %%%%%%%%%%%%
\begin{equation}
\label{pert}
\frac{\beta {\langle H' \rangle}_{H_0}}{N}=
\frac{3}{5} \left( \frac{l_B}{R_g} \right) \left( 1- \rho \right)^2
\left( 1-\frac{2R_g}{3R_{ws}} \right) \ ,
\end{equation}
where the terms  $ {\cal O} \left( R_g^3/R_{ws}^3 \right)$ are omitted.
The OCP-part is also easily computed; it is somewhat cumbersome to be
written explicitly in a general case. We analyze behavior of the system
in the limit $R_{ws} \gg R_g$ and  $N \gg 1$. Keeping for the free energy
only leading terms with respect to vanishing $ R_g/R_{ws}$ and
$1/N$, analyzing relative contribution of different addendums and
omitting less significant ones, we finally arrive at the following
result  for the total (conditional) free energy:
\begin{eqnarray}
\label{F_total1}
&&\frac{\beta F (\alpha, \rho)}{N} \simeq
\frac{9}{4N} \left\{ \alpha^2 + \frac{1}{\alpha^{2}}
  + \frac{{B}^* N^{1/2}}{ \alpha^{3}} \right\}
- 3 (1-\rho) \ln {R}^*_{ws} \nonumber \\
&&- \frac32 \left(\frac{2}{\pi^2} \right)^{1/3}
  \frac{{l}^*_B \rho^{4/3}}{N^{1/6}\alpha}
+ \frac35 \frac{{l}^*_B N^{1/2}}{\alpha} (1-\rho)^2 \ ,
\end{eqnarray}
where we introduced dimensionless ${l}^*_B\equiv l_B 6^{1/2}/a$ and
${R}^*_{ws} \equiv R_{ws}/a$.

The equilibrium state of the system is determined by the free energy
minimum with respect to both variables $\alpha$ and $\rho$.
Eq. (\ref{F_total1}) clearly demonstrates a competition for the equilibrium
``in-core'' counterion density  $\rho$  between the two largest
(at $\rho<1$) terms, second and fourth in the right-hand side of
Eq. (\ref{F_total1}). The second, negative term, large for $R_{ws} \gg 1$,
accounts for the counterions entropy. It tends to minimize the free energy
by minimizing $\rho$; i.e. it drives the counterions apart from the
polyelectrolyte in order to fill uniformly all the space of the
Wigner-Seitz cell. The positive fourth term (also large at $N \gg 1$)
accounts for the free energy of screened Coulombic interaction between
monomers. It is minimal if all the counterions are condensed on the
polymer, i.e. when $\rho =1$. Thus mainly this two-term competition
determines the equilibrium counterion density. The third term in the 
right-hand 
side of Eq. (\ref{F_total1}), which accounts for the counterion density
fluctuations, becomes important only if $\rho\rightarrow 1$. Next, the free
energy is to be minimized with respect to the expansion factor $\alpha$. At
this step, the first term in the right-hand side of Eq. (\ref{F_total1})
(which does not depend on $\rho$) becomes important. The following
asymptotic cases give explicit solutions:

(i) If ${l}^*_B N^{1/2} \ll \alpha \ln {R}^*_{ws}$, the second
negative counterion-distribution entropic term is the most important in Eq.
(\ref{F_total1}) and $\rho \rightarrow 0$ as $\sim (R_g/R_{ws})^3$. This
case corresponds to the unscreened Coulombic interactions between the chain
segments, so that polyelectrolyte tends to expand, and $\alpha \gg 1$. We
can write in the leading terms now
\begin{equation}
\label{F_total2}
\left. \frac{\beta F}{N} \right|_{\rho \to 0} \simeq
\frac{9\alpha^2}{4N} + \frac35 \frac{{l}^*_B N^{1/2}}{\alpha}
+\mbox{const} \ .
\end{equation}
Thus the equilibrium expansion factor and the radius of gyration are,
respectively,
\begin{equation}
\label{al1}
\alpha \simeq \left(\frac{2}{15}\right)^{1/3} {l}_B^{*\, 1/3} N^{1/2}
\qquad \mbox{or} \qquad
\frac{R_g}{a} \sim {l}_B^{*\, 1/3} N \ .
\end{equation}
This regime corresponds to the polyelectrolyte strong stretching.

(ii) If ${l}^*_B N^{1/2} \gg \alpha \ln {R}^*_{ws}$, the fourth
positive, inter-segment screened-Coulombic interaction term in
Eq. (\ref{F_total1}) is essentially larger than the counterion entropic
term, $\rho\simeq 1$ and $\alpha \ll 1$. Thus,
\begin{equation}
\label{F_total3}
\left. \frac{\beta F}{N} \right|_{\rho \to 1} \simeq
\frac{9{B}^*}{4N^{1/2}\alpha^3}
- \frac32 \left(\frac{2}{\pi^2} \right)^{1/3}
  \frac{{l}^*_B}{N^{1/6}\alpha}
\end{equation}
and the equilibrium dimensions are
\begin{equation}
\label{al2}
\alpha \simeq 3 \left(\frac{\pi}{4}\right)^{1/3}
\frac{{B}^{*\, 1/2}}{{l}_B^{*\, 1/2} N^{1/6}}
\qquad \mbox{or} \qquad
\frac{R_g}{a} \sim \frac{{B}^{*\, 1/2} N^{1/3}}{{l}_B^{*\, 1/2}} \ .
\end{equation}
This is a regime of the polyelectrolyte strong collapse.

For the general case we solved the minimization problem numerically (again
for the conditions, $R_{ws} \gg R_g$ and $N \gg 1$). We analyzed the
dependences of $R_g$ on ${l}_B$ at fixed ${B}^*$, ${R}^*_{ws}$ and
$N$. The $R_g({l}^*_B)$-dependence for some particular values of the
parameters are shown in Figure~\ref{fig:1}. For small and large values
 of the reduced
Bjerrum length, ${l}^*_B$, the radius $R_g$ is changing in accordance
with the asymptotic Eqs. (\ref{al1}) and (\ref{al2}). Our findings are in
qualitative agreement with the results of the numerical study in Ref.
\cite{Kremer1}, where the same two different regimes in the
$R_g({l}^*_B)$-dependence were observed. The most interesting is
however the case of the intermediate values of ${l}^*_B$. In this
region we observed a sharp bend in the dependence of the equilibrium free
energy on ${l}^*_B$ with a discontinuity of its first-order derivatives.
We interpret this as a first-order phase transition from the strong
stretching regime, with $\alpha \gg 1$, to the strong collapse regime, with
$\alpha \ll 1$. This phase transition is accompanied (or driven) by the
process of the counterion condensation, when the counterion density changes
from $\rho \ll 1$ (counterions are uniformly spread over the bulk) to
$\rho \simeq 1$ (practically all counterions are confined inside the
polyelectrolyte globule). After the counterion condensation the
polyelectrolyte dimensions become essentially smaller than they would be
for a neutral chain with the same volume interactions. 

\begin{minipage}{8cm}
\begin{figure}[htbp]
  \centerline{\psfig{file=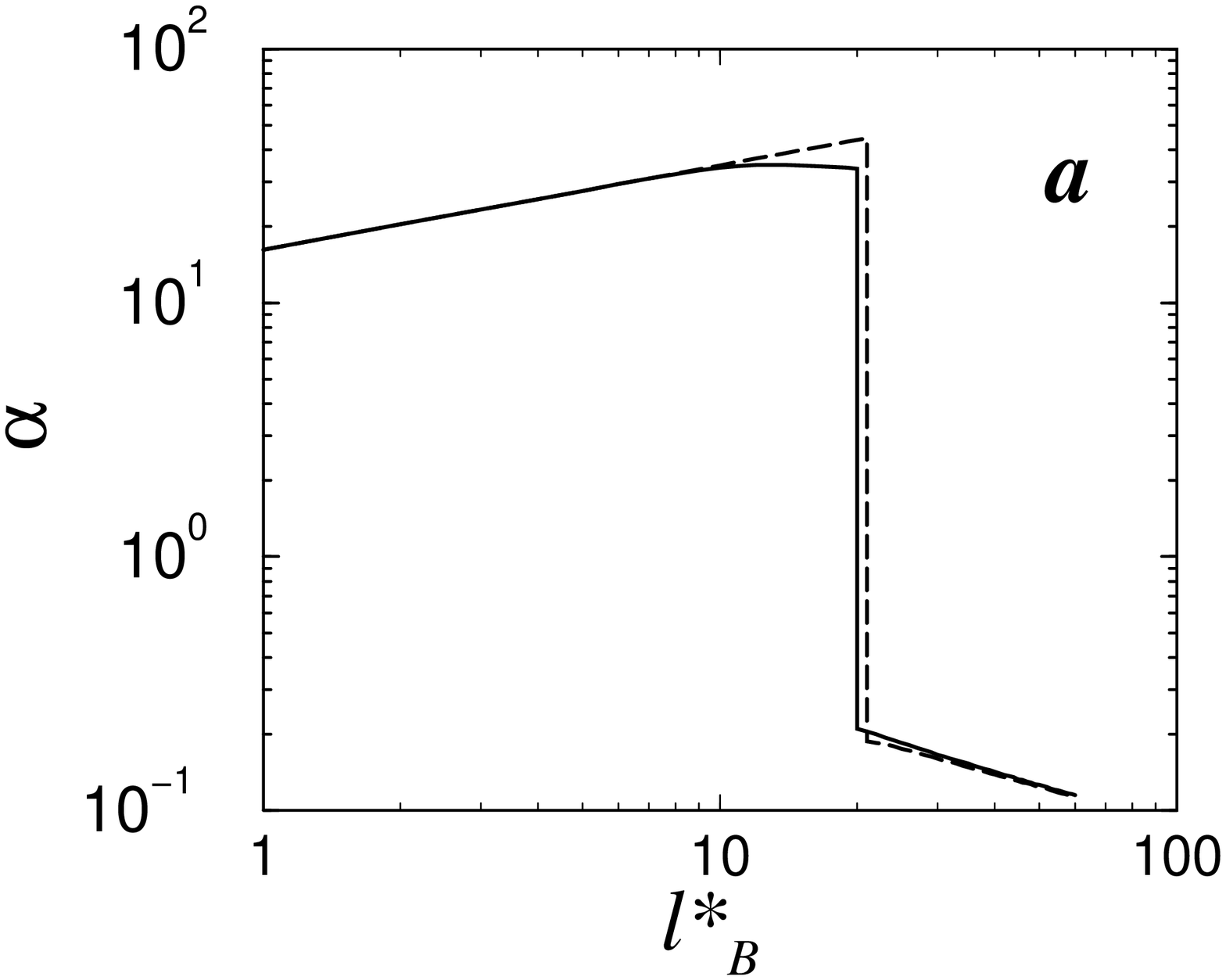,width=7.5cm}}
  \centerline{\psfig{file=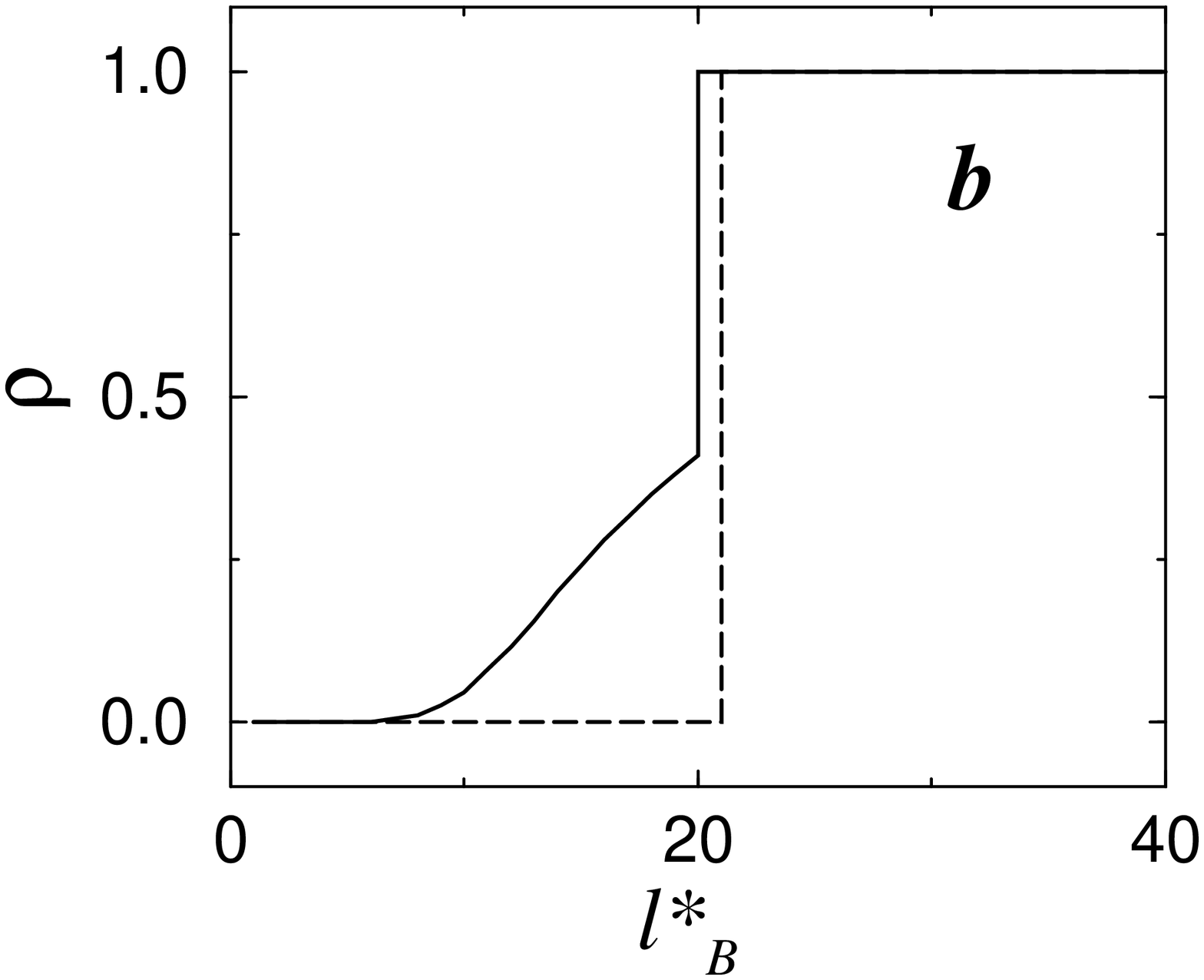,width=7.5cm}}
%,height=4.8cm}}
\vspace{1cm}
  \caption{(a) The logarithmic dependences of the polyelectrolyte
expansion factor $\alpha$ versus the reduced Bjerrum length,
${l}^*_B\equiv l_B \sqrt{6}/a$, calculated for $N=10^3$,
${R}^*_{ws}=10^5$ with ${B}^*=1$. A dramatic jump-like decrease of
the polyelectrolyte size occurs along with the first order phase transition
due to the counterion condensation on the
polyelectrolytes. (b) The corresponding dependences of the 
near-polyelectrolyte counterion density. Dotted curves refer to the 
numerical analysis of the simplified Eq.(\ref{F_total1}). Solid curves 
account for the next-order corrections with respect to the small 
parameters $\left(R_g/R_{ws} \right)$ and $\left( 1/N \right)$.}
  \label{fig:1}
\end{figure}
\end{minipage}

The qualitative
explanation to this phenomenon follows from the fact that for the strong
Coulombic interactions (${l}^*_B \gg 1$) the OCP-part of the free
energy gives rise to a  negative pressure which takes over the ideal chain
and counterions entropic pressure. It may be balanced by the intersegment
noncoulombic repulsion but only at some degree of compression.

In conclusion, we developed a simple theory of the linear polyelectrolyte
salt-free dilute solutions. We analyzed the dependence of the gyration
radius of the chain on the Bjerrum length, $l_B$, which characterizes the
strength of the Coulombic interactions in the system and found two
different regimes in this dependence, the same as were observed in
numerical studies. Additionally, we detected a first-order phase transition
from chain stretching to strong collapse, which is accompanied by the
counterion condensation on the polyelectrolyte.

Helpful discussions with B.Weyerich are highly appreciated.

% ***** ***** ***** ***** ***** ***** ***** ***** ***** *****
%\vfill\newpage

\vskip .8cm

\end{multicols}


\begin{thebibliography}
\bibitem{remark11}
In all expressions below we omit the ``self-energy'' parts $\propto l_BN
\varphi (0)$;  they have no impact on the  results obtained.
\end{thebibliography}
\end{document}